\documentclass{aa}

\usepackage{silence}
\usepackage{graphicx}
\usepackage{txfonts}
\usepackage{hyperref}
\usepackage{xcolor}
\usepackage{natbib}
\usepackage{upgreek}
\usepackage{mathtools}
\usepackage{amsmath}
\usepackage{listings}

\newcommand{\apt}{\textsc{AutoPhOT}\xspace}
\newcommand{\iraf}{\textsc{IRAF}\xspace}
\newcommand{\snoopy}{\textsc{SNOoPY}\xspace}
\newcommand{\astropy}{\textsc{ASTROPY}\xspace}
\newcommand{\pyraf}{\textsc{PyRAF}\xspace}
\newcommand{\py}{\textsc{Python}\xspace}
\newcommand{\hotpants}{\textsc{HOTPANTS}\xspace}
\newcommand{\pyzogy}{\textsc{PyZogy}\xspace}
\newcommand{\lmfit}{\textsc{LMFIT}\xspace}
\newcommand{\daophot}{\textsc{DAOPHOT}\xspace}

\newcommand{\pvb}{\textcolor{black}{Elias-Rosa et al. in prep}\xspace}
\newcommand{\wxt}{\textcolor{black}{Engrave Collaboration et al. in prep}\xspace}

\graphicspath{{./Figures/}}

\begin{document} 

\title{The AUTOmated Photometry Of Transients pipeline}

\subtitle{\apt}

\author{S. J. Brennan$^{1}$ and M. Fraser$^{1}$}
\authorrunning{S. J. Brennan \& M. Fraser}

\institute{$^{1}$ School of Physics, O’Brien Centre for Science North, University College Dublin, Belfield, Dublin 4, Ireland}

\date{Received XXX ; accepted XXX}

\abstract
 {We present the \textit{AUTOmated Photometry Of Transients} (\apt) package, a novel automated pipeline that is designed for rapid, publication-quality photometry of transients. \apt is built from the ground up using \py 3 - with no dependencies on legacy software. Capabilities of \apt include aperture and PSF-fitting photometry, template subtraction, and calculation of limiting magnitudes through artificial source injection. \apt is also capable of calibrating photometry against either survey catalogs (e.g. SDSS, PanSTARRS), or using a custom set of local photometric standards.
 We demonstrate the ability of \apt to reproduce lightcurves found in the published literature. \apt's ability to recover source fluxes is consistent with commonly used software e.g. \daophot, using both aperture and PSF photometry. We also demonstrate that \apt can reproduce published lightcurves for a selection of transients with minimal human intervention.} 


\maketitle


\section{Introduction}
For over three decades, the most commonly used packages for photometry are part of the Image Reduction and Analysis Facility (\iraf)\footnote{\iraf is distributed by the National Optical Astronomy Observatory, which is operated by the Association of Universities for Research in Astronomy (AURA) under cooperative agreement with the National Science Foundation} \citep{toby_iraf1986,toby_iraf1993}. Within \iraf, \textsc{ \daophot} \citep{DAOPHOT1987} is a suite of packages designed to perform photometry in crowded fields (i.e when sources are closely spaced together). 

In 2013, the National Optical Astronomy Observatories (NOAO) suspended further development of \iraf, and since then a community of astronomers has worked on maintaining the packages and adapting the current version (V2.16 / March 22, 2012) to work on modern hardware. However, a large portion of \iraf code cannot be compiled as a 64-bit executable, and must be built as a 32-bit program. Recently, several popular operating systems (e.g. MacOS) have dropped 32-bit support, which is required for \iraf. With continued development, as well as the emergence of new programming languages, \iraf has become more and more difficult to build and maintain on current architectures. Furthermore, \pyraf \citep{pyraf2012}, the main \py 2.7 wrapper for \iraf, has lost support and, as of January 1 2020, Users have been encourage to move to the currently supported \py 3 framework.

Besides \iraf /  \daophot there are a number of other photometry packages in use today. \textsc{SExtractor} \citep{Bertin1996} is a source detection and deblending tool used extensively for photometric measurements and is the basis for many modern photometric pipelines \citep[e.g. ][]{Mommert_2017,Merlin2019}. Other stand-alone photometry packages have been developed such as \textsc{A-PHOT} \citep{Merlin2019} and \textsc{PhotometryPipeline} \citep{Mommert_2017}, that mainly perform aperture photometry on ground based images.

Photometry tools have also been developed as part of \astropy \citep{astropy:2013,astropy:2018}, which is a community led project to develop a set of core software tools for astronomy in \py 3. 

In this paper we present the \textit{AUTOmated PHotometry of Transients} Pipeline (hereafter refereed to as \apt). \apt was designed to provide a fast, precise, and accurate means to measure the magnitude of astronomical point sources with little human interaction. The software has been built from the ground up, removing any dependence on the commonly used \iraf or any deprecated \py packages (for example those that rely on \py2).

\apt is designed to address some of the specific needs to astronomers working of transient phenomena such as supernovae. Observational campaigns for transients often yield heterogeneous datasets, which include images spanning several nights to decades, taken in a variety of photometric bands, and using different telescope and instrument configurations. For precise photometry, careful extraction of photometric data is required. However, the effect of different instruments and slightly different filter throughputs can increase the overall scatter in photometric data. Furthermore, photometry performed by different astronomers may show discrepancies based on the choice of parameters used e.g. quality/number of sequence stars used, aperture size, background subtraction etc.

\apt uses \astropy packages extensively. As \astropy is community driven, widely used, and written in \py 3, \apt is likely to have support from these packages for the foreseeable future. 

\apt can accept astronomical images from most ground based telescopes and cameras, and will adapt to image quality and/or telescope parameters to provide a homogeneous photometric output. \apt is available on \textsc{Github}\footnote{\url{https://github.com/Astro-Sean/autophot}} and available for installation though \textsc{Conda} \footnote{\url{https://anaconda.org/astro-sean/autophot}}. \apt will receive continued support, and one should refer to the online documentation for up-to-date information and further implementations \footnote{\url{https://autophot.readthedocs.io/en/latest/}}.

The purpose of this paper is to briefly outline the \apt package\footnote{Version 1.0.2}. We discuss the automated preprocessing within \apt in Sect. \ref{sec:pre_processing} and how photometric measurements are made in Sect. \ref{sec:photometry}. We provide a brief outline of the photometric calibration in Sect. \ref{sec:correction}. We outline the limiting magnitude package in Sect. \ref{sec:lim_mag}. Finally, we discuss the performance of \apt and its ability to provide science-ready results in Sect. \ref{sec:discussion}.

\section{Pre-processing}\label{sec:pre_processing}

\subsection{Image reduction}

Due to the specific nuances of various CCDs, it is left to the User or observing facility to correctly reduce the images prior to running \apt. These steps should typically include bias, flat-field and bad pixel corrections. For a general overview on these reduction steps, see \cite{howell_2006}. 

\subsection{Image stacking}

\apt does not perform image stacking. Often multiple exposures will be taken in the same bandpass during the night, in particular when long exposures that are susceptible to cosmic rays are used.

It is difficult to produce a universal image stacking procedure, and it is hence left to the User to stack images if they so wish\footnote{For example with \texttt{CCDPROC} \citep{matt_craig_2017}}. \apt hence treats multiple images taken on the same night independently. The User is cautioned that if they combine images, they should update the header keywords for gain and readout noise where necessary before running \apt.

\subsection{Target Identification}\label{sec:target_photometry}

\apt implements the Transient Name Server\footnote{\url{https://www.wis-tns.org/}} (TNS) \py API to obtain the most up-to-date coordinates of a particular transient. These coordinates are transformed from right ascension (RA) and declination (Dec) into X and Y pixel coordinates using the image World Coordinate System (WCS), see Sec. \ref{sec:astrometry}.

If a transient is not known to the TNS then the RA and DEC can be manually specifed by the User.

\subsection{Parsing image and instrument metadata}\label{sec:telescope_compatibility}

Flexible Image Transport System (\textit{FITS}) files are commonly used to store astronomical image. These files typically contain a 2D image as well as the image metadata stored as keyword-value pairs in a human-readable ASCII header. While \textit{FITS} header keywords contain critical information about the observation itself, such as exposure time, filter, telescope, these keywords are often inconsistent between different observatories.

When \apt is run on an image from a new telescope, the software asks the User to clarify certain keywords using the \texttt{check\_teledata} package. For example, this may involve clarifying whether ``\textit{SDSS-U}'' refers to Sloan $u$ or Johnson-Cousins $U$. This is the only step in running autophot which requires human intervention, but is necessary due to the ambiguous filter naming conventions used by some telescopes.

After the \apt telescope check function has run, the results are saved as a human-readable \textit{Yaml} file (see example in Listing \ref{code:telescope.yml}) allowing for easy additions, alterations or corrections. When \apt is subsequently run on images from the same telescope and instrument, it will lookup filter names etc. in this \textit{Yaml} file.

\begin{lstlisting}[language=python,caption={Example of entry in \texttt{telescope.yml} for the Nordic Optical Telescope (NOT). This entry includes instrument-specific information needed for header keyword translation (FILTER, AIRMASS, GAIN), filter keywords (g\_SDSS: g, B\_Bes: B, etc) as well as location information and extinction terms, discussed further in Sect. \ref{sec:airmass_correction}},label=code:telescope.yml,float]
NOT:
 INSTRUME:
    ALFOSC_FASU:
        Name: NOT+ALFOSC
        AIRMASS: AIRMASS
        GAIN: GAIN
        RDNOISE: READNOISE
        filter_key_0: FILTER
        filter_key_1: FILTER1
        pixel_scale: 0.213
        B_Bes: B
        V_Bes: V
        color_index:
            B:
                B-V:
                    m: 0.014
                    m_err: 0.007
            V:
                B-V:
                    m: -0.106
                    m_err: 0.012
            .
            .
            .
    NOTCAM:
        Name: NOT+NOTCAM
        AIRMASS: AIRMASS
        GAIN: GAIN
        .
        .
        .
 
 extinction:
    ex_B: 0.203
    ex_I: 0.019
    ex_R: 0.069
    .
    .
    .
     
 location:
    alt: 2327
    lat: 28.76
    lon: -17.88
    name: lapalma

\end{lstlisting}

Along with filter names, the \texttt{Yaml} database (shown in Listing \ref{code:telescope.yml}) contains other instrument-specific information necessary for automated execution of \apt. The nested dictionary structure allows for multiple instruments at the same telescope (in the example shown information is given for both the ALFOSC and NOTCam instruments mounted on the Nordic Optical Telescope).

\textit{filter\_key\_0} gives the fits header key which gives the filter names\footnote{At the time of writing these filters are Johnson-Cousins $U$, $B$, $V$, $R$, $I$, Sloan- $u$, $g$, $r$, $i$, $z$ and NIR- $J$, $H$, $K$ }. To account for instruments with multiple filter wheels, this keyword can be iterated i.e. filter\_key\_0, filter\_key\_1, etc. If it finds an incompatible header value i.e. if the filter corresponds to $CLEAR$ or $AIR$, it is ignored unless requested otherwise by the User\footnote{This is necessary in the case of unfiltered observations.}. 

\apt requires at minimum for an image to have the \textit{TELESCOPE} and \textit{INSTRUME} keywords. Both keywords are standard fits keywords\footnote{\url{https://heasarc.gsfc.nasa.gov/docs/fcg/standard_dict.html}} and are virtually ubiquitous across all astronomical images. If not found, an error is raised and the User is asked for their intervention.

A pre-populated \textit{Yaml} file with information and keywords for several commonly-used telescopes is provided as part of \apt .

\subsection{Solving for the World Coordinate System}\label{sec:astrometry}

Astronomical images require a World Coordinate System (WCS) to convert sky coordinates to X and Y pixel coordinates. Many images may have WCS values written during the reduction process. However, it is not uncommon for an image to have an offset WCS or be missing WCS information entirely. \apt assumes the WCS is unreliable when there is a significant (default is $2\times FWHM$) offset between the catalog positions of sources in the image, and their measured position.
In such cases (and where a WCS is missing entirely), \apt calls a local instance of \textsc{Astrometry.net}\footnote{\url{https://astrometry.net/use.html}} \citep{Lang_2010}. Source detection is performed on the input image, and asterisms (sets of four or five stars) are geometrically matched to pre-indexed catalogs. Solving for the WCS values typically takes from $\sim5s$ to $\sim30s$ per image\footnote{Using a 2017 MacBook Pro, with a 2.5 GHz Intel Core i7 processor with 8 Gb DDR3 RAM.}. 

\subsection{Cosmic Ray removal}\label{sec:cray}

Cosmic Rays (CRs) are high energy particles that impact the CCD detector and can result in bright points or streaks on the CCD image. For images with long exposure times, CRs can be problematic as they may lie on top of regions or sources of interest.

To mask and remove cosmic rays, \apt uses an instance of \texttt{Astroscrappy}\footnote{\url{https://github.com/astropy/astroscrappy}} \citep{vanDokkum_2012,astroscrappy_2019} which is a \py 3 adaptation of the commonly used \texttt{LACosmic} code \citep{vanDokkum_2012}.

\subsection{Measuring image Full Width Half Maximum}\label{sec:FWHM}

The Full Width Half Maximum (FWHM) of point sources in an image is determined by the astronomical seeing when the image was taken, as well as the telescope and instrument optics. \apt measures the FWHM of an image by fitting an analytical model (by default a Moffat function; \citealp{Moffat_1969}) to a few tens of bright isolated sources in the field. 

\begin{figure}
\centering
\includegraphics[width= \columnwidth]{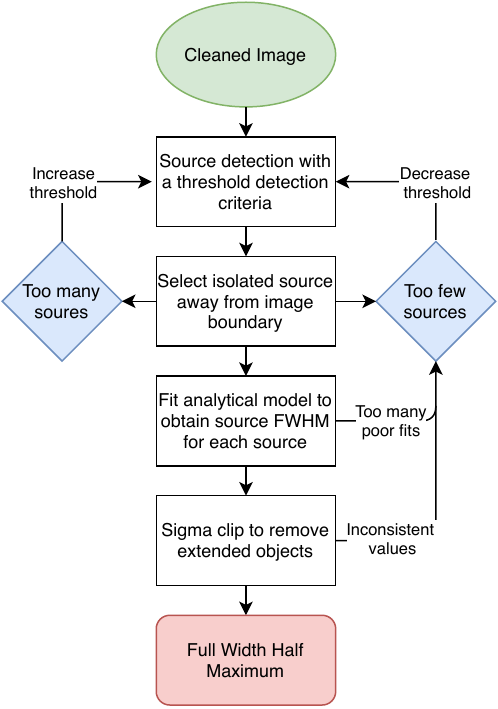}
\caption{Flowchart showing the iterative process of finding the FWHM of image using \apt. We do not show the adaptive threshold step size for purpose of clarity.}
\label{fig:FWHM_flowchart}
\end{figure}

Firstly, \apt needs to adapt to the number of point sources in an image. A deep image with a large field of view (FoV) will have considerably more sources than a shallow image with a small FoV. Too few sources may lead to poorly sampled (or incorrect) values of the FWHM, while too many sources may indicate the detection threshold is too low (i.e. background noise is detected as a source) and needlessly increases the computation time.
Fig. \ref{fig:FWHM_flowchart} illustrates the process for finding the FWHM of an image. \apt's FWHM function in the \texttt{FIND} package aims to obtain a well sampled value for the FWHM of the image without any prior knowledge of the number of sources in the field. The process begins with a search for point-like searches using the \texttt{DAOFIND} \citep{DAOPHOT1987} algorithm, together with an initial guess for the threshold value (that is, the minimum counts above the background level for a source to be considered). The first iteration returns a small set of bright sources, measures their FWHM and updates the initial guess for the FWHM value. 

The process continues to search for sources above the threshold value in the field. If too many sources are detected, the loop will rerun the algorithm with a higher threshold value. This change in threshold value is adaptively set based on the number of sources detected. 

We use sigma clipping is used to remove extended sources (e.g. faint galaxies, saturated sources, etc) which may have slipped through. In classical sigma clipping, if we have a median value for the FWHM with a standard deviation, $\sigma$, then only values with within $\pm n \sigma$ of the median is used, where n is some value, which by default is set to $n=3$. \apt uses a more robust method to determine outliers via the median absolute deviation  given by:

\begin{equation}\label{eq:MAD_std}
\begin{split}
\sigma_{MAD} = \frac{MAD}{\Phi^{-1}(P)} \approx 1.4826 \cdot MAD \\
 \rm{where}~MAD = median(| X_i - \mu |)
\end{split}
\end{equation}

\noindent where $\Phi^{-1}(P)$ is the normal inverse cumulative distribution function evaluated at probability P = 3/4. Assuming a normal distribution of FWHM values, $n=3$ would mean that $\sim99\%$ of FWHM measurement would fall within this value. Once a FWHM value is found for an image it is then used henceforth for this image for building the PSF model and photometric measurements.

\section{Photometry}\label{sec:photometry}

Fundamentally, photometry consists of the measuring the incident photon flux from an astronomical source and calibrating this onto a standard system. We can define the difference in magnitude between two sources $m_1$ and $m_2$ as

\begin{equation}\label{eq:magnitude_difference}
 m_1 - m_2 = -2.5 \cdot Log_{10}\left(\frac{F_1}{F_2}\right)
\end{equation}

\noindent where $F_1$ and $F_2$ are the measured fluxes (counts per second) from two sources. As Eq. \ref{eq:magnitude_difference} describes a relative system, we also need to define some fiducial stars with known magnitudes. One such definition is the ``Vega'' magnitude system, where the magnitude of the star Vega in any given filter is taken to be 0\footnote{In practice, the modern definition of the Vega magnitude system implies that Vega itself has a magnitude of 0.03.} 
In this case, the magnitude of any other star is simply related to the flux ratio of that star and Vega as follows:

\begin{equation}\label{eq:apparent_magnitude_vega}
 m_1 = -2.5 \cdot Log_{10}\left(\frac{F_1}{F_{Vega}}\right)
\end{equation}

When performing photometry on transients, we typically measure the instrumental magnitude of the transient itself as well as several reference sources with known catalog magnitudes in the image. Comparing the magnitude offset with the literature values of these reference sources (which can be unique to each image due to varying nightly conditions) and applying it to the transient, we can place the measurement of the transient onto a standard system. We define the apparent magnitude of the transient as

\begin{equation}\label{eq:apparent_magnitude_image}
 \begin{split}
 m_T = -2.5 \cdot Log_{10}(F_T) + \left\langle \sum_i m_{cat,i} + 2.5 \cdot Log_{10}(F_i)\right\rangle \\
 \rightarrow m_T = m_{instrumental,T} + ZP
 \end{split}
\end{equation}

\noindent where $m_T$ is the unknown apparent magnitude of the transient with a flux $F_T$. The later term describes the magnitude offset or zeropoint (ZP) for the image and is found by subtracting the catalog magnitude , $m_{cat,i}$, from the measured magnitude, $-2.5 \cdot log_{10}(F_i)$. An average value for the zeropoint is typically calculated using a few tens of sources in the field, typically close to the transient position. 

Applying simply a zeropoint will typically result in photometry that is accurate to $O\sim0.1$ mag or better. For more precise calibration, and in particular to ensure homogeneous measurements across different instruments, one must apply additional corrections beside the zeropoint. These include color correction ($CC_\lambda$) terms and aperture corrections, which we discuss in Sect. \ref{sec:correction}.

\subsection{Aperture Photometry}\label{sec:AP_phot}

\apt can perform either aperture or PSF-fitting photometry on images. Both methods have their advantages and limitations. Aperture photometry is a simple way to measure the number of counts within a defined area around a source. This technique makes no assumption about the shape of the source and simply involves summing up the counts within a aperture of a certain radius placed at the centroid of a source. 

\apt begins by using aperture photometry as an initial guess to find the approximate magnitude of bright sources. If PSF-fitting photometry is not used, for example if it fails due to a lack of bright isolated sources in the field\footnote{Sources must have a signal to noise ratio of greater that 25 to be used in the PSF model by default.}, aperture photometry is implemented. Aperture photometry can yield accurate results for bright, isolated sources (flux dominated), but may give measurements with larger uncertainties for faint sources (noise dominated), see Appendix \ref{app:error}.

To perform aperture photometry, \apt first finds the centroid of a point source by fitting a Gaussian function. To accurately measure the brightness of a source, the background flux must be subtracted. This can be done in several ways in \apt, including a local median subtraction or fitting a 2D polynomial surface to the background region.

Choosing the optimum background subtraction requires some prior knowledge of the FWHM. The median subtraction method is best for a cutout with a flat background (e.g. template subtracted images, see Sect. \ref{sec:templates}), or for a smoothly varying background over the scale of a few FWHM. For a background with strong variations (e.g. a on the edge of a extended source) the surface fitting algorithm performs best. For consistency, \apt retains the same background subtraction method (surface fitting by default) for all point source measurements. 

\begin{figure}
\centering
\includegraphics[width= \columnwidth]{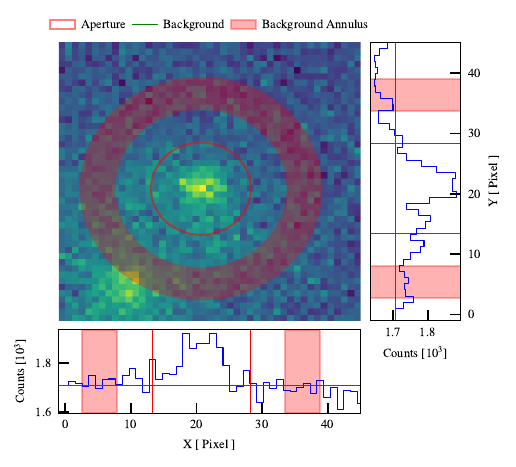}
\caption{Aperture photometry of an isolated point source showing the aperture radius (solid red line) as well as the background annulus (faded regions) and assumed background value (green solid line, taken as the median annulus value). We also include projections along the X and Y axes. }
\label{fig:ap_phot}
\end{figure}

We demonstrate the aperture photometry functionality in Fig. \ref{fig:ap_phot}. In this case the background counts have been found within an annulus centred on the source position. The counts from the source can then be found using:

\begin{equation}\label{eq:aperture_phot}
\begin{split}
F_\lambda \times t_{exp} = counts_{ap} - \langle counts_{sky} \rangle \cdot n_{pix}\\
\end{split}
\end{equation}

\noindent where $counts_{ap}$ is the total counts within the aperture area, $\langle counts_{sky} \rangle$ is the average counts due to the sky background and $n_{pix}$ is the number of pixels within our aperture.

\begin{figure}
\centering
\includegraphics[width= \columnwidth]{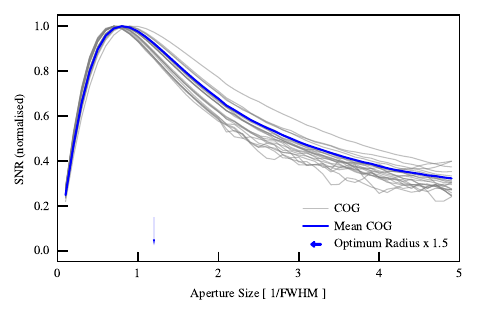}
\caption{Sample of S/N values within a given aperture given in the major axis. Mean Curve Of Growth (COG) given as blue and optimum radius taken as maximum of this curve, given by blue arrow. The optimum radius is multiplied by 1.5 to allow for any discrepancy in centroiding.}
\label{fig:COG}
\end{figure}

There is a balance when selecting an optimum aperture size. The aperture should be large such that most of the light from the star is captured. However, it should be small enough so that contamination from the sky background and unrelated sources in minimized. Fig. \ref{fig:COG} demonstrates a search for optimum aperture size in \apt\footnote{This value can also be fixed and the default is taken as 1.6 $\times$ FWHM}. For a sample of bright sources found from Sect. \ref{sec:FWHM}, the signal-to-noise ratio is measured within a series of apertures of increasing radii. Typically the S/N will reach a maximum at 1--2 times the FWHM, although this can vary depending on the PSF. The aperture radius at which the S/N is maximized is then multiplied by 1.5 to allow for any error in centroiding and used as new aperture radius for the image.

To account for any discrepancy in aperture size (e.g. missing flux due to finite aperture size) we employ an aperture correction to account for noise dominated wings of faint sources, which will miss counts due to lower S/N of the PSF wings. A smaller aperture will lead to a larger aperture correction and vice-versa, with typical corrections being less $\sim0.1$~mag. This is not necessary if PSF photometry is used.

\begin{figure}
\centering
\includegraphics[width= \columnwidth]{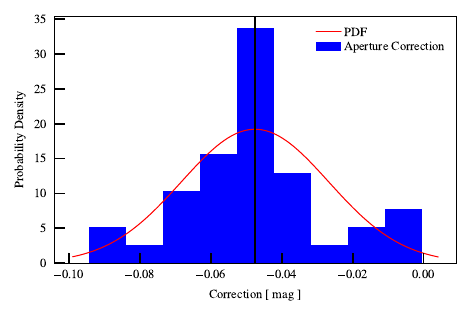}
\caption{ Histogram shows magnitude of the ratio of our large aperture size with $r=2.5\times FWHM$ and normal aperture size with $r=1.6\times FWHM$ for a single image. This is the aperture correction used when aperture photometry is employed.}
\label{fig:aperture_correction}
\end{figure}

To calculate the aperture correction, the bright sources found in Sect. \ref{sec:FWHM} are measured with a \textit{large aperture size} and the standard aperture size. \apt uses an large aperture size with $r = 2.5 \times FWHM$ and normal aperture size with $r=1.6 \times FWHM$\footnote{Both values are adjusted if an optimum aperture size is used}. Using Eq. \ref{eq:magnitude_difference}, the ratio of these values gives a magnitude correction which compensates for the flux lost due to a finite aperture size. In Fig. \ref{fig:aperture_correction} we plot the distribution of aperture corrections for a sample of bright, isolated sources. The average value and standard deviation are taken as the aperture correction, which is applied to all sources measured with standard aperture size during aperture photometry. 

Aperture photometry has its drawbacks. It performs poorly in crowded fields, where contamination from neighbouring sources can interfere with measurements of a single point source. Additionally transients that occur close to their host may have complex backgrounds which may contaminate measurements. Aperture photometry is more susceptible to CCD detector defects such as hot/cold and dead pixels and CRs. Moreover, aperture photometry assumes a flat weight function across the aperture and is susceptible to centroiding discrepancies. Point sources inherently have a Gaussian-like weight function which can more accurately account for the PSF. 

Although aperture photometry can always be used (with varying results), modelling the PSF of a star can provide more accurate measurements and can be applied to more dynamic scenarios, such as blended sources and high background scenarios.

\subsection{Point Spread Function Photometry}\label{sec:PSF_phot}

All point sources in an image, regardless of their brightness or colour have the same \textit{shape} to their PSF\footnote{As long as the sources are unresolved and not saturated.}. The amplitude of the PSF will of course change with brightness.

PSF-fitting photometry uses bright sources in the field to build a semi-analytical model which is then fitted to fainter sources to find their instrumental magnitude. PSF photometry is the method of choice for crowded fields and can give better results for low S/N sources when compared to aperture photometry. 

\apt assumes that the PSF is non-spatially varying across the image, meaning points sources will in theory appear the same regardless of their location on the image. In practice this may not be the case for images that cover a large FoV \citep{howell_2006}. If \apt detects a significant variation in PSF shape across the images, if will only perform measurements within a radius around the transient position where the PSF is approximately constant.

The PSF package designed for \apt is based on the work of \cite{DAOPHOT1987}, \cite{Massey1992} and \cite{Heasley1999}. \apt uses ``well-behaved'' sources to build the PSF model which will then be used to measure the amplitude of sources in the field. These sources must be have a high S/N, isolated from their neighbours and have a relatively smooth background. This is done by building a compound model comprised of an analytical component (such as Gaussian or Moffat) along with a numerical residual table obtained during the fitting process. Although sources are selected from Sect. \ref{sec:FWHM}, the User may supply the coordinates of PSF stars.

If a FWHM of an image is comparable to the pixel size, the image is said to be under sampled. In this case PSF-fitting photometry is particularly susceptible to centroiding errors \citep{Wildey_1992,Lauer99}. If \apt finds a very small FWHM for an image (default is 2 pixels) aperture photometry is used instead.

\begin{figure*}
\centering
\includegraphics[width= \textwidth]{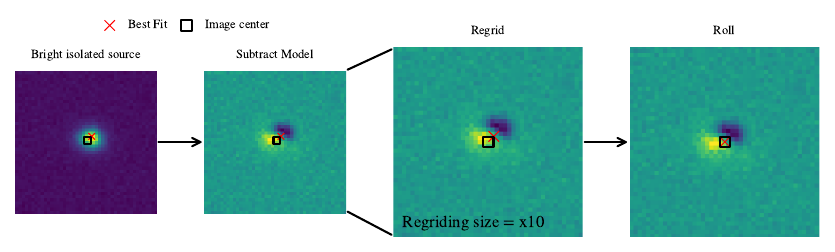}
\caption{Demonstration of the steps taken to build residual table for PSF photometry. Bright isolated sources are located as in Sect. \ref{sec:FWHM}. A cutout is taken around the source, and an analytical function is fitted and subtracted. The image is then resampled to a finer pixel grid (default: $\times$10). The residual image is then rolled (discretely shifted along x and y) such that the location of best fit is at the image center. This is repeated for several bright isolated sources to obtain an average residual. This figure can be produced in \apt using the \texttt{plots\_PSF\_model\_residual} command.}
\label{fig:PSF_residualtab}
\end{figure*}

Figure \ref{fig:PSF_residualtab} illustrates the process of building a PSF model in \apt. Bright isolated sources are located and fitted with an analytical function (first panel). The best fit location is noted and the analytic model is subtracted to leave a residual image (second panel). The residual image is resampled onto a finer pixel grid and shifted (third panel and fourth panel). The compound (analytic and residual) PSF model is then normalized to unity. This process is repeated for several (typically $\sim$ 10 sources) bright isolated sources, to create an average residual image. The final step is to resample the average residual image back to to the original pixel scale. We ensure flux in conserved during this process. Our final PSF model is then simply:

\begin{equation}\label{eq:PSF_equation}
PSF(x,y,A) = M(x_0,y_0,A,FWHM) + R(x_0,y_0,A)
\end{equation}

\noindent where $M$ is the a 2D Moffat function (or Gaussian function if selected) and $R$ is the residual image. We can fix the FWHM to the value found for the image as discussed in Sect. \ref{sec:FWHM}, so the PSF model can be fitted with three parameters, $x_0$ and $y_0$ (the centroid of the sources), and $A$ its amplitude. 

We integrate under the analytical model between bounds set by the FWHM and aperture size, and perform aperture photometry on the residual table with the same bounds. This is the counts in a PSF model with amplitude equal to 1. When fitting the PSF model we implement the same re-sampling technique to allow for sub-pixel fitting. We can then simple multiply the fitted amplitude of the source with the counts under our normalised PSF model to find the counts for any given source.

\begin{figure*}
\centering
\includegraphics[width= \textwidth]{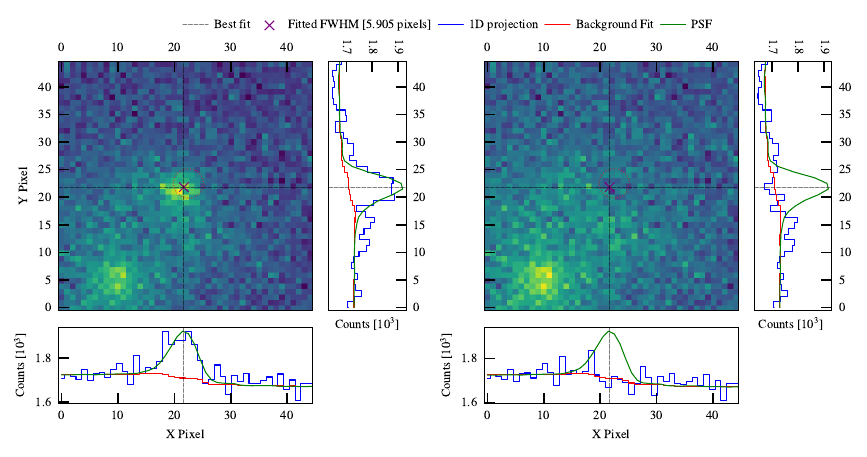}
\caption{Example of PSF subtraction. The main panels show a cutout of the transient location before (left) and after (right) PSF subtraction, while projections along the x and y axis are also shown for each panel. The source is cleanly subtracted and there is no sign of a residual in the subtracted panel. This figure is can be reproduced using the \texttt{plot\_PSF\_residuals} command. }
\label{fig:PSF_subtraction}
\end{figure*}

In Fig. \ref{fig:PSF_subtraction} we show an example of the residual image after fitting our PSF model to a source and subtracting it off. In this example the point source is almost symmetric. A signature of a suitable PSF model is that after subtraction, there is little to no evidence of the prior point source.

\section{Calibrating photometry}\label{sec:correction}

A crucial step in photometry is calibrating instrumental magnitudes onto a standard photometric system. Due to the sparsity of photometric nights (i.e. nights when there are no clouds or other issues with atmospheric transparency), this zeropoint calibration must be obtained for each image. Furthermore, even on photometric nights, there may be a gradual shift in zeropoint due to the cleanliness/coating of the mirrors over time \citep[for an example of this effect, see fig. 3 in][]{harbeck2018}. We discuss the zeropoint calibration in Sect. \ref{sec:zeropoint}. In some cases it is sufficient to apply the zeropoint correction alone to produce calibrated, publication-ready photometry.

However, in cases where multiple instruments have been used to observe a supernova measurements, one must account for differences between telescopes. In particular, we must consider effects due slight manufacturing differences between filter sets which may give systematic offsets for the same transients measured using different instruments. These effects typically accounts for $\sim0.1~mag$ corrections to photometry.

\subsection{Zeropoint Calibration}\label{sec:zeropoint}

The zeropoint is used to calibrate an instrumental magnitude to a standard magnitude system using  Eq. \ref{eq:apparent_magnitude_image}.
For a given image, \apt will search for a catalog covering the field of interest with the correct filters available. Alternatively, the User can specify their desired catalog, or provide a custom catalog for the field. 

\begin{figure}
\centering
\includegraphics[width= \columnwidth]{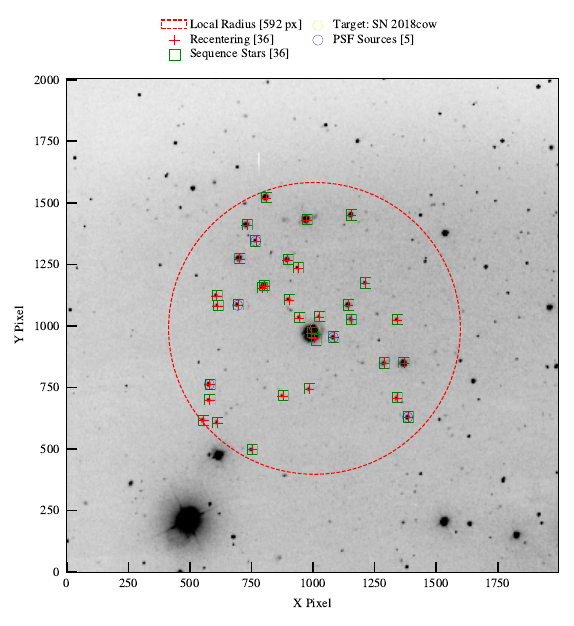}
\caption{Demonstration of source detection for catalog sources. PSF stars (blue circles) are selected on the basis of their brightness and isolation. In this example we only consider sources close to the transient location ($<3'$)
}
\label{fig:source_selection}
\end{figure}

Fig. \ref{fig:source_selection} illustrates how sources are identified in an image to determine the zeropoint as well to build the PSF model. In this example a local region  3\arcmin\ around the target position is selected. If the image contained many sources, this can reduce computation times considerably.

\begin{figure}
\centering
\includegraphics[width= \columnwidth]{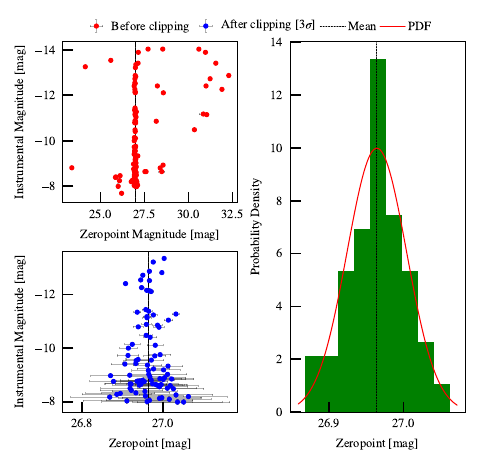}
\caption{Zeropoint diagnostic plot from \apt. Left panels show zeropoint measurements before (upper left) and after (lower left) a 3$\sigma$ clipping. Upper left panel shows a skewed tail for fainter instrumental magnitudes (notice the different y-axis between the two subplots). Right panel shows zeropoint distribution with a probability density function with a well defined peak. }
\label{fig:zeropoint}
\end{figure}

We show the zeropoint calibration for the image shown in \ref{fig:source_selection} in Fig. \ref{fig:zeropoint}. In this example we include sigma-clipping (see Sect. \ref{sec:FWHM}) to remove any outliers as well as a S/N cutoff. The result shows a distribution with a well defined peak which is used as the zeropoint for this image. 

\subsection{Color terms}\label{sec:color_correction}

Along with the zeropoint, it is usually necessary to apply colour terms when calibrating instrumental magnitudes. Colour terms are a consequence of filters and CCDs having a non-uniform response over the bandpass of a filter. For example, a $z$-band filter may transmit light with wavelengths between 8200 and 9200~\AA. However, if this filter is used with a CCD that has a much lower quantum efficiency in the red, then we will detect more counts from a blue source than a red, {\it even if they have the same z-band magnitude}. This effect, which manifests itself as a colour-dependent shift in zeropoint, can be as much as 0.1 mag. Moreover due to small differences in the effective pass band of different observatory filter system, we must determine the color term for each instrument individually to a produce a homogeneous dataset.

\begin{figure}
\centering
\includegraphics[width= \columnwidth]{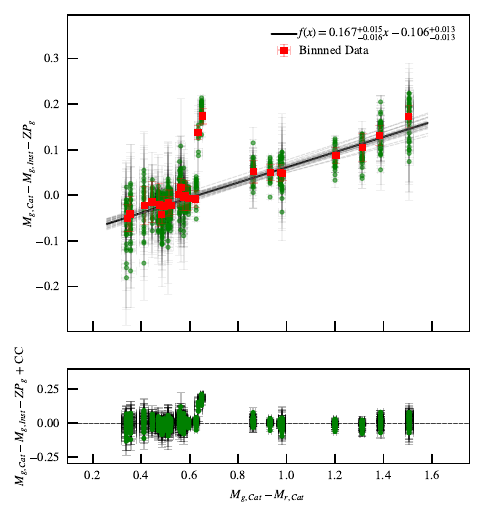}
\caption{Demonstration of the effect of point source colour on zeropoint calibration. X-axis shows the catalog colour of sources, while the Y-axis shows the g-band magnitudes minus their instrumental magnitude and image zeropoint.  Red squares are binned magnitudes with errorbars equal to the standard deviation of magnitudes in each bin. The solid black line shows the best fit using EMCEE \citep{Foreman_Mackey_2013}. The lower panel shows the same points with the color correction applied. }
\label{fig:color_MCMC_fit}
\end{figure}

\begin{figure}
\centering
\includegraphics[width= \columnwidth]{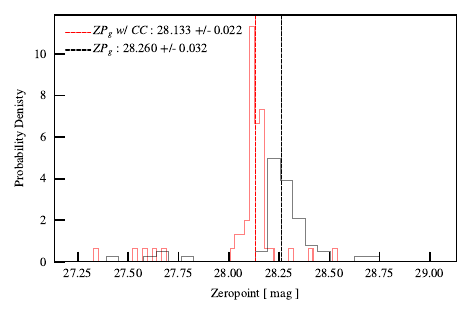}
\caption{Discrepancy of zeropoint measurements when ignoring color correction (black) and including it (red).}
\label{fig:corrected_zeropoint}
\end{figure}

We demonstrate the effect of neglecting any colour information when determining the zeropoint of an image in Fig. \ref{fig:color_MCMC_fit} and Fig. \ref{fig:corrected_zeropoint}. A clear discrepancy is seen and is correlated with the color of the sequence stars used; in this case, the zeropoint under represents blue sources and slightly overestimates redder sources by $\sim0.1$-mag. In Fig. \ref{fig:corrected_zeropoint}, we see a shift of $\sim0.1$-mag in the zeropoint magnitude as well as smaller scatter among sources in the field.

For transient measurements, observations in two closely spaced filters are required e.g. B-V, taken at approximately the same time. Additionally the color term of the instrument and telescope must be known. This can be found using stars in the field with standard magnitudes in literature to determine the effect of stellar color i.e. the fitted line given in Fig. \ref{fig:color_MCMC_fit}. The slope of this line ($CT_{BV}$) is then used to correct for the zeropoint for each image where appropriate color information is available. As we have more unknowns variables than known, we can iterate through Eq. \ref{eq:color_iteration} to solve for the true, color corrected magnitude.

\begin{equation}\label{eq:color_iteration}
\begin{aligned}
 M_{True,~B,~i} = m_{inst,~B} + ZP_{B} + CT_{B,BV}~(M_{True,~B,~i-1}-M_{True,~V,~i-1})\\ 
 M_{True,~V,~i} = m_{inst,~V} + ZP_{V} + CT_{V,VB}~(M_{True,~B,~i-1}-M_{True,~V,~i-1})
\end{aligned}
\end{equation}

\noindent The above equation demonstrates the process of applying a colour correction to two measurements in filters $B$ and $V$. Both filters have a colour term known a priori, where $CT_{B,BV}$ is the slope of $M_B$ - $M_V$ v.s. $M_B$ - $M_{B,inst}$ and similarly for $CT_{B,BA}$. For convenience and stability, \apt solves for the color term corrections using the iterative Jacobi method. We rearrange Eq. \ref{eq:color_iteration} into the form $Ax = b$ which gives:

\begin{gather}\label{eq:color_Abx}
\begin{bmatrix} 1 - CT_{B,BV} & CT_{B,BV} \\ -CT_{V,VB} & 1+CT_{V,VB} \end{bmatrix}
\begin{bmatrix}
M_{True,~B,~i}\\ 
M_{True,~V,~i}
\end{bmatrix}
=
\begin{bmatrix}
m_{inst,~B} + ZP_{B}\\ 
m_{inst,~V} + ZP_{V} 
 \end{bmatrix}
\end{gather}

\noindent This is a quick method to apply a colour correction and typically converges in $\sim$10 iterations. 

\subsection{Atmospheric Extinction}\label{sec:airmass_correction}

We can account for the effect of atmospheric extinction using the following:

\begin{equation}\label{eq:airmass_extinction}
\begin{aligned}
M_{\lambda,corrected} = M_{\lambda} + \kappa_\lambda \cdot sec(z)
\end{aligned}
\end{equation}

\noindent where $M_{\lambda}$ is the magnitude in a given filter, $\lambda$, $\kappa_\lambda$ is the extinction coefficient in magnitudes per unit airmass and $sec(z)$ is simply the secant of the zenith angle $z$. Taking account of the airmass correction is particularly necessary when calibrating photometry to standard fields \citep[e.g.][]{Landolt1992}. An observer may wish to obtain a more precise set of sequence stars for their transient measurements. This will involve observing a standard field on a night that is photometric, as well as the transient location. The zeropoint measurements of the standard field will be at a different airmass than the transient. Using Eq. \ref{eq:airmass_extinction}, and the standard field measurement, an observer can perform photometry on a set of sequence stars around the transient location and place them on a standard system. This can be used for future measurements of the transient.

There is no trivial way to approximate the extinction at a specific telescope site. We provide an approximation which \apt uses in Appendix \ref{app:extinction}, although for accurate photometry, the User should provide the known extinction curve for a given site.

\section{Image Subtraction}\label{sec:templates}

If a transient is close to its host nucleus, occurs near another point source, or has faded to a level comparable to the background, it may be necessary to perform difference imaging \citep[e.g.][]{alard1998}. Difference imaging involves scaling and subtracting a template images (assumed to have no transient flux) from a science images, removing a strong bright or host contamination. Prior to subtraction images must be precisely aligned (i.e. to subpixel precision), scaled to a common intensity, and be convolved with a kernel so that their PSFs match. 

Currently, \apt includes \hotpants \footnote{High Order Transform of PSF ANd Template Subtraction}\footnote{\url{https://github.com/acbecker/hotpants}} \citep{HOTPANTS_2015} and \pyzogy \footnote{\url{https://github.com/dguevel/PyZOGY}} \citep{Zackay_2016} for image subtraction. The User can select what package they require, with \hotpants set as the default. Prior to template subtraction, \apt aligns the science and template images using WCS alignment\footnote{\url{https://reproject.readthedocs.io/en/stable/}} or point source alignment\footnote{\url{https://astroalign.readthedocs.io/en/latest/}}\citep{Beroiz_2019}. Furthermore both images are cropped to exclude any regions with no flux after alignment.

\section{Limiting Magnitude}\label{sec:lim_mag}

A limiting magnitude is the brightest magnitude a source could have and remain undetected at a given significance level. Even when a transient is not visible in an image, a limiting magnitude can help constrain explosion times in SNe or decay rates of GW merger events.

\begin{figure*}
\centering
\includegraphics[width= \textwidth]{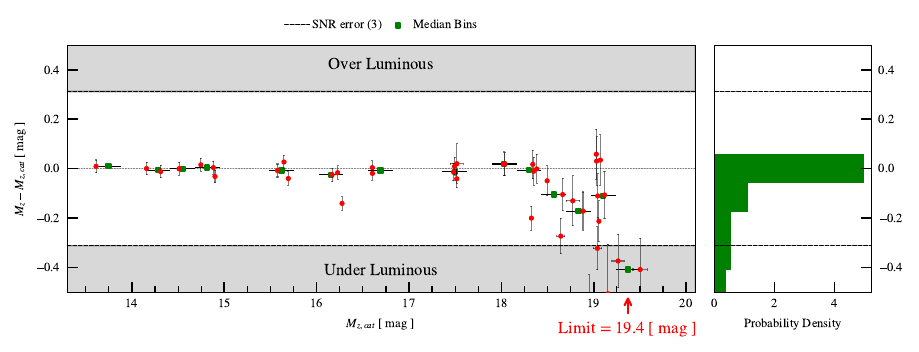}
\caption{Difference between measured magnitude versus catalog magnitude for sequence sources in an image. We adopt a 3$\sigma$ threshold for source detection corresponding to $\delta(M)\approx0.31~mag$ using Eq. \ref{eq:S/N_err}.}
\label{fig:catalog_detection}
\end{figure*}

A relatively crude method to find the limiting magnitude of an image is to attempt to recover known sources in the FoV. In Fig. \ref{fig:catalog_detection} we show the difference between the recovered magnitude of sources in an image and their catalog magnitude. The difference is close to zero for the majority of brighter sources, but then becomes significant for fainter sources with $M_{Catalog} > 18.1$. \apt will calculate the limiting magnitude from the first magnitude bin where the difference exceeds a specified threshold (set using the equations in Appendix \ref{app:error}). Although this is relatively straightforward, it can fail in sparse fields, and of course is not feasible when the image is deeper than the catalog.

\begin{figure}
\centering
\includegraphics[width= \columnwidth]{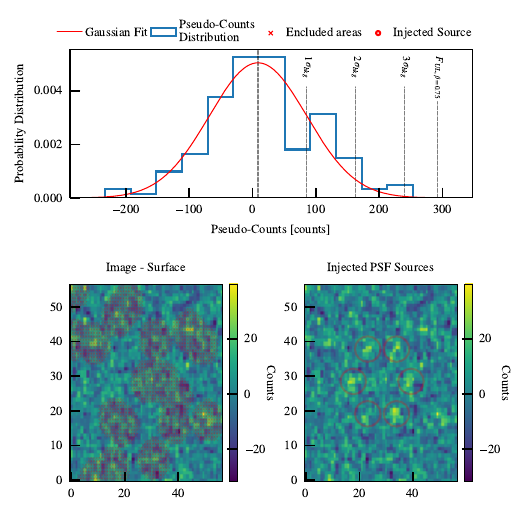}
\caption{The upper panel shows the distribution of summed counts for a random set of pixels close to the expected source location. Bottom left shows a cutout of the transient location; pixels marked in red are excluded when creating the distribution. Bottom right is the same image with injected PSF sources (marked with red circles) with magnitude equal to the $F_{UL,\beta=0.75}$ limiting magnitude,see Appendix. \ref{app:compute_FU}}
\label{fig:prob_limiting_mag}
\end{figure}

The second way that \apt can calculate the limiting magnitude is through what we refer to as the ``probabilistic limiting magnitude'' illustrated in Fig. \ref{fig:prob_limiting_mag}. We assume that the pixels are uncorrelated, and contain only noise from a uniform background sky.  After excising the expected position of the transient, we proceed to select $n$ pixels at random (where $n=\pi r^2$), and sum together the counts in these $n$ pixels from a background subtracted cutout of the transient location. Repeating this many times for different random sets of $n$ pixels, we obtain a distribution of summed counts (shown in the upper panel in Fig. \ref{fig:prob_limiting_mag}). We can then ask the question ``what is the probability we would obtain this number of counts or greater by chance?''. Setting the threshold to $3\sigma$, in the example shown we can see that we are unlikely to find a source with more than $\sim3,500$ counts, and we hence adopt this as our limiting magnitude.

Finally, the most rigorous limiting magnitude is determined though injecting and recovering artificial sources. Using an initial guess from the probabilistic limiting magnitude described above, artificial sources built from the PSF model (see Sect.\ref{sec:PSF_phot}) and with realistic noise are injected in set positions (default $3~\times~FWHM$) around the target location. The magnitudes of the injected sources are then gradually adjusted until they are  no longer recovered by \apt above $3\sigma$ (or some other criteria).

\begin{figure}
\centering
\includegraphics[width= \columnwidth]{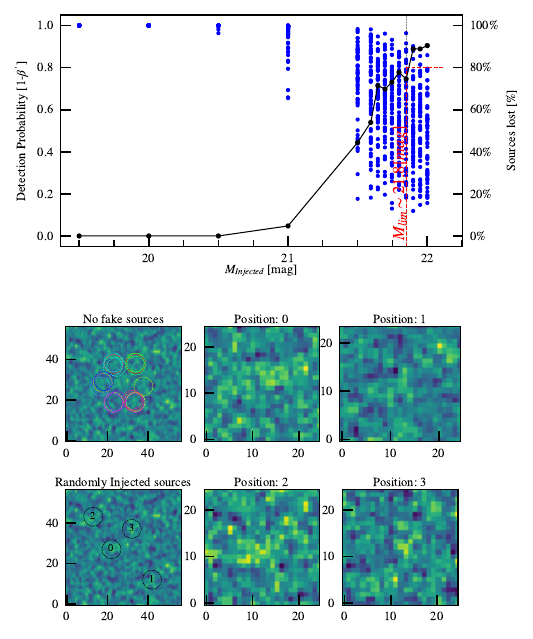}
\caption{Diagnostic plot from \apt's artificial source injection package. Top panel shows the change in the detection probability ($1-\beta'$) for artificially injected sources. In this example the sources are considered lost at $\beta=0.75$ and the detection cutoff is reached when $80\%$ of sources are lost (black line with circles). The leftmost image cutouts illustrate locations around the target location before (upper) and after (lower) sources were injected randomly at the limiting magnitude. The remaining four panels demonstrate closes up of these injected sources}
\label{fig:inject_limiting_mag}
\end{figure}

Fig. \ref{fig:inject_limiting_mag} demonstrates the artificial source injection package. In this example the image is template subtracted and we use the $\beta'$ detection criteria (see Appendix \ref{app:compute_FU}). Starting with an initial guess from the probabilistic limiting magnitude, the injected magnitude is adjusted incrementally until it meets our detection criteria, which it will typically overshoot. The magnitude increment is then reversed, using a smaller step size until the detection criteria is again fulfilled. Sources are deemed lost when where their individual recovered measurements give a $\beta<0.75$. We take the limiting magnitude to be the magnitude at which 80\% of sources are lost.

\section{Testing and validation}\label{sec:discussion}

\subsection{Testing of photometry packages}\label{sec:testing_phot}

In this section we demonstrate \apt's ability to recover the magnitude of sequence stars in the field. As this is a novel PSF-fitting package, we compare against the aperture photometry package available in \apt as well as from the well established photometry package \daophot \citep{DAOPHOT1987}.

\begin{figure}
\centering
\includegraphics[width= \columnwidth]{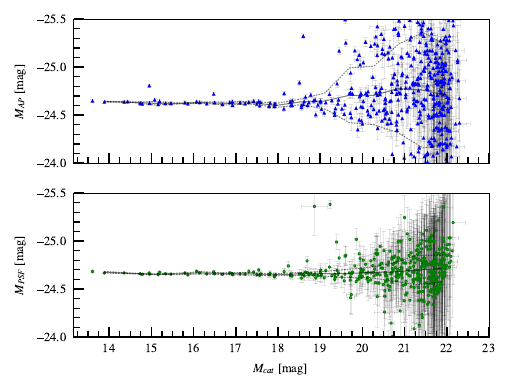}
\caption{ Demonstration of recovered magnitude using aperture photometry (upper panel) and the PSF-fitting package (lower panel) from \apt. Y-axis shoes the derived zeropoint magnitude for each source. Solid lines show a moving mean value with dashed lines indicating the standard deviation in each bin. Horizontal errorbars show the uncertainty on catalog magnitudes, vertical errorbars are uncertainties on recovered magnitudes from \apt. }
\label{fig:psf_vs_aperture}
\end{figure}

Fig. \ref{fig:psf_vs_aperture} shows both aperture and PSF photometry can accurately determine the magnitude of relatively bright sources ($M\lessapprox19~mag$). However, at fainter magnitudes, aperture photometry can longer performs as well, as seen in the larger scatter. Incorrect centroiding may become an issue with aperture photometry when the source flux is comparable to the background. PSF photometry can perform much better at fainter magnitudes. Unlike aperture photometry, the PSF model attempts to measure shape of a point-like source using more information on the shape of the PSF.  

\begin{figure}
\centering
\includegraphics[width= \columnwidth]{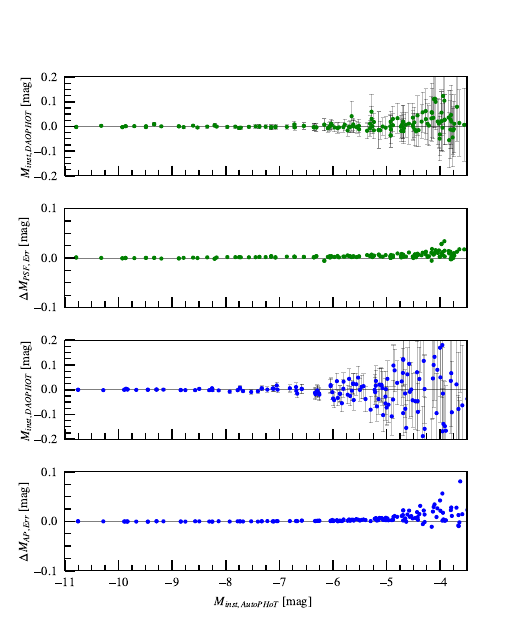}
\caption{Comparison of measured instrumental magnitude using \daophot and \apt. The upper two panels show the difference in recovered magnitude using PSF-fitting photometry with \apt and \daophot, and the difference in error. The lower two panels show the same but for aperture photometry. In each case the x-axis gives the instrumental magnitude from \apt. The same aperture radius was used in all cases. For the first and third panel, the errorbars are the combination of (added in quadrature) uncertainties from both \apt and \daophot. In panels two and four, the y axis shows \daophot$_{err}$ - \apt$_{err}$ (i.e. the uncertainties from \apt are slightly smaller than those returned by \daophot for faint sources).}
\label{fig:DAOPHOT_comparision}
\end{figure}

Fig. \ref{fig:DAOPHOT_comparision} compares the PSF and aperture photometry from \apt and \daophot \citep{DAOPHOT1987}. The PSF fitting package from \apt can match the recovered instrumental magnitude from \daophot, even at faint magnitudes where the flux from the source becomes comparable to the sky background. Aperture photometry can result in similar magnitudes but suffers from centroiding errors at fainter magnitudes. However for such low fluxes, PSF-fitting photometry should be used. 

\begin{figure}
\centering
\includegraphics[width= \columnwidth]{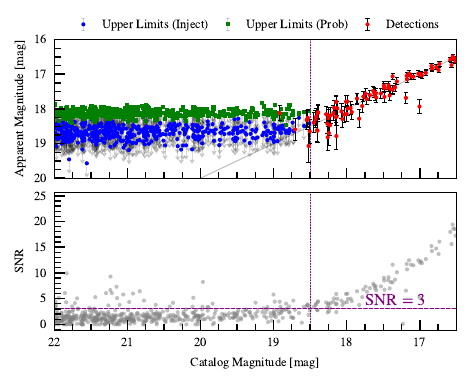}
\caption{Demonstration of \texttt{inject\_sources} function. Upper panel: if the recovered magnitude is less than the magnitude limit, the source cannot be recovered. Blue points indicate the magnitude of injected sources. In cases where the source could not be recovered we indicate the limiting magnitude with a green arrow, otherwise we plot recovered sources as red points. The solid line shows a perfect recovery of sources. Lower panel: Measured S/N for each source. The horizontal line highlights a S/N of 3 while the vertical line highlights the approximate magnitude of sources recovered, set at $18.5~mag$.}
\label{fig:detection_recovery}
\end{figure}

We test the effectiveness of the \apt limiting magnitude packages in Fig. \ref{fig:detection_recovery}. We use a relatively shallow image, and a reference catalog containing fainter sources. We see that below the computed upper limit of $\sim18.5~mag$, sources are not detected. Brighter than $\sim18.5~mag$, we recover sources at magnitudes consistent with their catalog values.

\subsection{Performance}\label{sec:performace}

\begin{figure*}
\centering
\includegraphics[width= \textwidth]{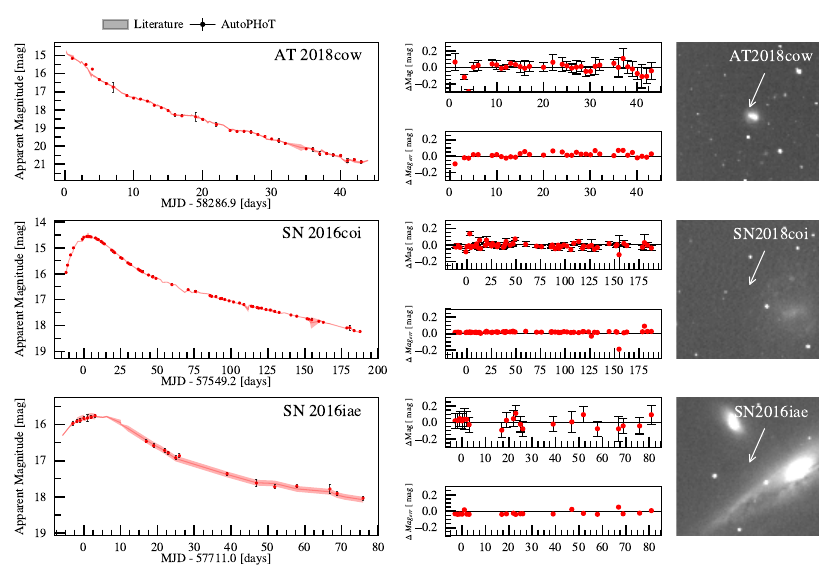}
\caption{r-band light curves produced by \apt compared to those found in literature. \apt points are given as points with error bars and literature values given as shaded band with width equal to the error at each point. In descending order, we compare the output from \apt for AT 2018cow \citep{perley_2018,Prentice_2018b} using template subtraction and aperture photometry, SN 2016coi \citep{Prentice_2018a} and SN~2016iae \citep{Prentice_2017} without subtraction using both PSF and aperture photometry. Center panels compare both measurements with the upper panel showing the difference between measurements ($\Delta~Mag = \apt - literature$) and lower panel showing the difference in error ($\Delta~Mag_{err} = \apt_{err} - literature_{err}$) for each transient. Right panels highlight the site of the transient event. }
\label{fig:performence}
\end{figure*}


Fig. \ref{fig:performence} shows a comparison of \apt photometry against published lightcurves in the literature for three transients found in three different environments, namely AT 2018cow \citep{perley_2018,Prentice_2018b}, SN 2016coi \citep{Prentice_2018a} and 2016iae \citep{Prentice_2017}. \apt was run on the same data as used in the referenced publications, and while a combination of techniques was used for each transient (i.e. template subtraction, PSF-fitting and aperture photometry) as detailed in the caption, in all cases this was run without human intervention.

We report several diagnostic parameters for these three transients in Tab. \ref{tab:performance}, including execution time. The most time consuming step is matching and fitting sequence stars to determine the zeropoint. This can be addressed by limiting the region where sequence sources are measured or providing \apt with a list of sources to use.

\begin{table*}
\begin{tabular}{c|c|c|c|c|c}
Transient  & Images & Time taken {[$hr$]} & Time per image {[$\frac{s}{image}$]} & Mean Residual {[$mag$]} &
Mean Error Residual {[}mag{]} \\
\hline 
AT 2018cow & 187     & 3.32                & 64                           & 0.041                      & 0.008                            \\
SN 2016iae & 259     & 4.69                & 65                           & 0.007                      & 0.020                            \\
SN 2018coi & 25      & 0.35                & 34                           & 0.011                      & 0.039                           
\end{tabular}
\caption{Performance of \apt computing $r$-band lightcurves from Fig. \ref{fig:performence}.
Photometry for AT 2018cow was performed using template subtraction and aperture photometry (similar to the \citealt{perley_2018}, although a custom host subtraction pipeline was used in this case.), whereas SN 2016coi \citep{Prentice_2018a} and 2016iae \citep{Prentice_2017} were reduced without subtraction using both PSF and aperture photometry where appropriate. Photometry performed using a 2017 MacBook Pro, using a 2.5 GHz Intel Core i7 processor with 8 Gb DDR3 RAM.}
\label{tab:performance}
\end{table*}

\section{Conclusions and Future Development}\label{sec:conclusion}

We present our photometry pipeline, Automated Photometry of Transients (\apt), a new publicly available code for performing PSF-fitting, aperture and template-subtraction photometry on astronomical images, as well as photometric calibration. This code is based on \py 3 and associated packages such as \astropy. With the deprecation of \py2 and popular photometry packages within \iraf, \apt provides accurate photometry with little User setup or monitoring.  \apt has already been used in several scientific publications (\citealt{chen_2021},  \citealt{Fraser_2021}, \citealt{brennan_2021a}, \citealt{brennan_2021b}, \pvb, and \wxt) at the time of writing.

Future work includes adapting to a wider range of images with irregularities, such as satellite trails, saturated sources, and CCD imperfections. The \apt project will also ultimately include a User-friendly web interface as well as an Application Programming Interface (API). This will allow for both fast and simple photometry without the need to maintain local software, as well as easy command line access. Additional functionality will allow for  calibrated photometry using standard fields observations. Further releases of \apt will include additional corrections such as spatially varying PSF models and potentially S-corrections \citep{Stritzinger2002}.

The pipeline is publicly available and detailed installation and execution instructions can be found from \url{https://github.com/Astro-Sean/autophot} 

\begin{appendix}

\section{Atmospheric Extinction Calculation}\label{app:extinction}

To deduce the extinction parameters across a range of photometric filters, one may observe a series of stars throughout a night at different airmasses. Fitting a slope to the data should show a clear trend of zeropoint magnitude versus airmass, being more extreme in the bluer bands than in the red. \apt can accept these values as shown in Code. \ref{code:telescope.yml}.

If unknown, \apt makes an rough approximation of the extinction due to airmass that relies on the altitude of the telescope site and the wavelength being observed. There are three main contributors to atmospheric extinction; Rayleigh scattering of light by molecules smaller than the wavelength of the scattered light, absorption due to Ozone in the upper atmosphere, and aerosol extinction by scattering and absorption by particles with diameters of the order of the wavelength or larger e.g dust/ash particles.

Absorption/scattering by Rayleigh scattering is described by \cite{Hayes1975} and is given by:

\begin{align*}\label{eq:kappa_rayleigh}
 & \alpha_{\lambda,Rayleigh} = 0.0094977 \cdot \lambda^{-4}\cdot n_s(\lambda)^2 \cdot e^{\frac{-h}{7.996}} \\ 
 & where~n_s(\lambda) = 0.23465 + \frac{107.6}{146-\lambda^{-2}} + \frac{0.93161}{41 - \lambda^{-2}},
\end{align*}

\noindent where $\lambda$ is the effective wavelength of the observation in $\mu$m and $h$ is the altitude above sea level in $km$. $n_s(\lambda)$ describes the refractive index of thin incoming light.

Molecular absorption, mainly due to atmospheric ozone and water, can be described by:

\begin{equation}\label{eq:kappa_ozone}
 \begin{aligned}
 \alpha_{\lambda,Ozone} = 1.11 \cdot T_{Ozone} \cdot \kappa_{Ozone}(\lambda)
 \end{aligned}
\end{equation}

\noindent where $T_{Ozone}$ is the thickness of the Ozone layer above the telescope scope taken at $0~^{\circ}C$ and 1~atm, and is assumed to be 0.3~cm. $\kappa_{Ozone}(\lambda)$ is the absorption coefficient for Ozone taken from \cite{Inn_53}.

By default, \apt will assume the total atmospheric extinction is $ \alpha_{\lambda} = \alpha_{\lambda,Rayleigh} + \alpha_{\lambda,Ozone}$. This can be a suitable approximation for many telescope sites e.g. La Silla and Roque de los Muchachos, Fig. \ref{fig:extinction}. In practice high particulate levels in the air can account for large discrepancies, especially in the redder bands, e.g. Paranal and Mauna Kea. This can be accounted for by including the atmospheric extinction due to aerosols. This is given by:

\begin{figure}
\centering
\includegraphics[width= \columnwidth]{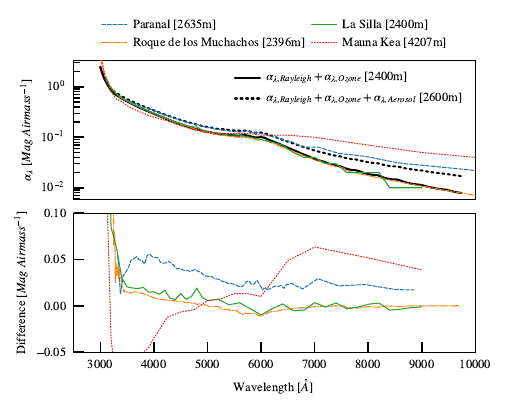}
\caption{Theoretical atmospheric extinction curves for several sites including La Silla \protect\footnote{https://www.eso.org/sci/observing/tools/Extinction.html}, Roque de los Muchachos\protect\footnote{https://www.ing.iac.es/astronomy/observing/conditions/wlext.dat}, Paranal\protect\citep{Patat_2011}, and Muana Kea\protect\footnote{https://www2.keck.hawaii.edu/inst/common/exts.html}. We include a match to the Paranal extinction curve using $\alpha_{\lambda} = \alpha_{\lambda,Rayleigh} + \alpha_{\lambda,Ozone}+\alpha_{\lambda,Aerosol}$ with b = -2, $A_0$ = 0.05 and $H_0$ = 1.5. It is difficult to fit the extinction curve at found at Muana Kea, likely due to high levels of volcanic dust.}
\label{fig:extinction}
\end{figure}

\begin{equation}\label{eq:kappa_aersol}
 \begin{aligned}
 \alpha_{\lambda,Aerosol} = A_0 \cdot \lambda^{-b} \cdot e^{\frac{-h}{H_o}}
 \end{aligned}
\end{equation}

\noindent where $A_0$ is the same extinction for $\lambda$=1~$\mu$m and $b$ is a coefficient dependent on the size of aerosol particles and their size distribution and $H_0$ is the scale height. The aerosol extinction is the most variable and problematic. It is left to the User to include this correction however with a priori knowledge, it can produce good results (e.g. for Paranel in Fig. \ref{fig:extinction}).

\section{Error Calculations}\label{app:error}

The uncertainty on the calibrated magnitude of a source is calculated as:

\begin{align}\label{eq:alinerror_eq}
 \delta m = \sqrt{ \delta m_{inst}^2 + \delta ZP^2 }
\end{align}

We take the error from the zeropoint calibration ($\delta ZP$) to be the standard deviation from measurements of sources in the field. Prior to this, appropriate sigma clipping and S/N cutoffs are applied. The error associated with the measurement of the transient itself ($\delta m_{inst}$) requires more attention.  The uncertainty in magnitude of a source is related to the S/N ratio as follows:

\begin{align}
m_{inst} \pm \delta~m_{inst} & = -2.5Log_{10} (S \pm N) \nonumber \\
 & = -2.5Log_{10}(S(1 \pm \frac{N}{S})) \nonumber \\
 & = -2.5Log_{10}(S) - 2.5Log_{10}(1+\frac{N}{S})\nonumber \\
 & = -2.5Log_{10}(S) - 2.5Log_{10}(1+\frac{1}{S/N})\nonumber\\
\delta~m_{inst} & = \mp 2.5Log_{10}(1+\frac{1}{S/N}) \approx 1.0875 (\frac{1}{S/N})
\label{eq:S/N_err}
\end{align}

\noindent Where $S$ is the signal from the source and $N$ is the noise associated with it. We find the the error associated with the S/N is $\sim 1.0875(\frac{1}{S/N})$ using a Taylor expansion. In \apt, we define the Signal to Noise Ratio (S/N) using the CCD equation \citep{Mortara_1981,howell_2006}:

{\small
\begin{align} 
S/N & = \nonumber \\
 & \frac{\overbracket{{F_* \times t_{exp}}}^{Signal}} {\sqrt{[ \underbracket{ (F_*\times  t_{exp})}_{Shot \ noise} + \underbracket{(F_{sky} \times t_{exp} \times n)}_{Sky \ Noise} + \underbracket{(RN^{2} + (G^2/4) \times n )}_{Read~Noise} + \underbracket{(D \times n \times t_{exp})}_{ Dark \ Current \ noise}]}}
\label{eq:S/N}
\end{align}
}%
 
\noindent here $F_*$ is the count rate from the star in $e^{-}/s$, $t_{exp}$ is the exposure time in seconds, $F_{sky}$ is the background counts in $e^{-}/s/pixel$, $n$ is the number of pixels within an aperture, $R$ is the read noise $e^{-}$, D is the dark current in $e^{-}/s$  and G is the Gain in $e^{-}$. R, G, $t_{exp}$, and D are taken from the image header, if available while the remaining terms are calculated during the photometric reduction.

Additionally we must consider the error associated with the fitting process itself. If PSF photometry is performed, we include an error estimate from artificial star experiments similar to those in \snoopy. If the User desires this additional error analysis, an artificial source with the same magnitude as the target star, is placed in the PSF-subtracted residual image in a position close to the real source (e.g. lower left panel of Fig. \ref{fig:inject_limiting_mag}). The injected sources are then recovered using an identical fitting procedure. The standard deviation of measurements is taken as an estimate of the instrumental magnitude error. This is combined (in quadrature) with the PSF-fit error returned by \lmfit to give  $\delta~m_{inst}$.

At the time of writing, \apt is only concerned with these terms given in Eq. \ref{eq:alinerror_eq} as these terms are expected to dominate.

\section{Execution Example}\label{app:execution}

In listing. \ref{code:execution_example} we provide a snippet of code that will execute \apt on a dataset\footnote{Example of \apt's execution can be found at \url{https://github.com/Astro-Sean/autophot}}. 

\begin{lstlisting}[language=python,caption={Example of \apt execution used to produce the lightcurve for SN 2016iae in Fig. \ref{fig:performence}.},label=code:execution_example,float]

# Import AutoPhoT package
import autophot

# Load command dictionary
from autophot.prep_input import load
autophot_input = load()

# location of work directory
autophot_input['wdir'] = '/Users/seanbrennan/Desktop/autophot_db'

# Location of fits images for SN2016iae
autophot_input['fits_dir'] = '/Users/seanbrennan/Desktop/SN2016iae'

# IAU name of target for TNS retrieval 
autophot_input['target_name'] = '2016iae'

# Name of catalog for zeropoint calibration
autophot_input['catalog']['use_catalog'] = 'apass'

# Import automatic photometry package
from autophot.autophot_main import run_automatic_autophot

# Run automatic phototmetry with input dictionary
run_automatic_autophot(autophot_input)

\end{lstlisting}

\section{Computing flux upper limits}\label{app:compute_FU}

As a transient fades to a magnitude which is comparable to the background brightness, it is necessary to compute detection criteria to determine whether a measured flux can be confidently associated with the transient.
Detection significance is usually defined in terms of the maximum probability of a false positive (a spurious detection of background noise), which we define as $\alpha$. Alpha can also be related to $\sigma$, e.g. a 3$\sigma$ upper limit will correspond to a 0.135\% probability (i.e. $\alpha$) of a false positive.


As part of \apt, we include a false negative criteria, $\beta$, which signifies the fraction of real sources that go undetected. This $\beta$ value can be defined in terms of a flux upper limit, $f_{UL}$, which  indicates the maximum incompleteness of a sample of sources with $f_{source} = f_{UL}$. In other words, $100(1 - \beta)\%$ of the sources with flux $f_{UL}$ will have flux measurements with a $S/N >n\sigma$. We follow the discussion of  $F_{UL}$ in \cite{Masci2011ComputingFU} and further detailed in \cite{Kashyap_2010}. We describe the probability of detection as $\beta' = 1-\beta$, which we want to maximise.  The probability of detecting a source with a $f_{source}=f_{ul}$ can be written as:

\begin{equation}\label{eq:probability_detection}
\begin{aligned}
\beta' =  \frac{1}{2}[1 - \textit{erf}(\frac{z}{\sqrt{2}})]\\
\rm{where}~z = \frac{n \sigma_{bkg} - f_{UL}}{\sigma_{bkg}}
\end{aligned}
\end{equation}

\noindent where $\textit{erf}$ is the error function, $n$ is a set detection limit (default to $3$ in \apt) and $f_{UL}$ is a flux upper-limit. Rearranging Eq. \ref{eq:probability_detection} gives probabilistic criteria for detection limits:

\begin{equation} 
\label{eq:upper_limit}
f_{UL} = [n +\sqrt{2}erf^{-1}(2\beta'-1) ]\sigma_{bkg}
\end{equation}

\noindent Using Eq. \ref{eq:upper_limit}, we see that using the common $f_{UL} = 3 \sigma_{bkg}$ gives a $\beta$' of 50\%, which means that roughly half of the sources injected at $f_{source}= 3\sigma_{bkg}$ will go undetected. For 95\% percent confidence that a source is genuine for S/N = 3, Eq. \ref{eq:upper_limit} gives an value for $ f_{UL} \approx 5\sigma_{bkg}$. 

\begin{figure}[!ht]
\centering
\includegraphics[width= \columnwidth]{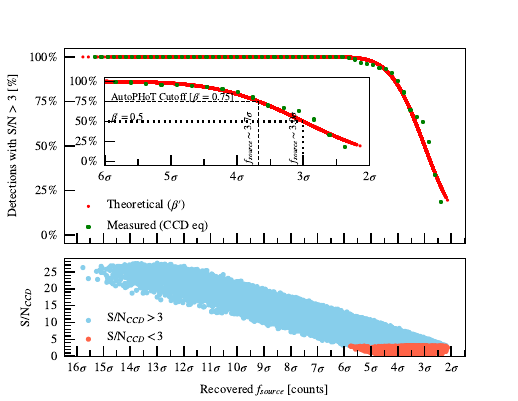}
\caption{Demonstration of Eq. \ref{eq:probability_detection}. We perform artificial source injection on an empty patch of sky. Sources are injected uniformly throughout the image, including sub-pixel placements, with random poisson noise added to the PSF prior to injection. In the upper panel, we plot Eq. \ref{eq:probability_detection} versus maximum pixel flux in units of the standard deviation of the background noise, as the red curve. The green points show the binned S/N ratio using Eq. \ref{eq:S/N}. In the lower panel we plot the S/N ratio using Eq. \ref{eq:S/N} with the same x-axis as the upper panel. The points are coloured blue if S/N > 3 and red if S/N < 3. For the sample of sources incrementally injected, Eq. \ref{eq:probability_detection} can reproduce the recovered fraction of sources. In other words, for sources measured with a $f_{source} \approx 3\sigma_{bkg}$, roughly half of these are recovered with a S/N > 3; for injected sources with $f_{source} \approx 3.7\sigma_{bkg}$ we detect roughly $75\%$; while virtually all sources are confidently recovered at $f_{source} \approx 4.5\sigma_{bkg}$ }
\label{fig:beta_criteria}
\end{figure}

We demonstrate the applicability of this $\beta$ criteria in Fig. \ref{fig:beta_criteria}. We inject artificial sources in an empty part of the sky, and while noting the injected source parameters, we assess whether or not the sources can be recovered to an appropriated S/N, see caption of Fig. \ref{fig:beta_criteria} for further details.

\apt defaults to $\beta'=0.75$ to provide a conservative upper limit to any reported limiting magnitudes. The User may also opt for a more traditional detection criteria of using the $S/N > n$ where $n$ is the level above $\sigma_{bkg}$.

\end{appendix}

\section*{Acknowledgements}

SJB would like to acknowledge the support of Science Foundation Ireland and the Royal Society (RS-EA/3471). MF is supported by a Royal Society - Science Foundation Ireland University Research Fellowship. We thank Simon Prentice for sharing data for AT 2018cow, SN 2016coi, and SN 2016iae, and comments on \apt during its early phase of development. We thank Kim Phan for correcting an error with the color term correct. We thank Emma Callis, Robert Byrne, Beth Fitzpatrick, Shane Moran, Kate Maguire,  and Llu\'is Galbany for their comments and feedback while \apt was under development. We thank the Transient Name Server and their open source sample codes\footnote{\url{https://www.wis-tns.org/content/tns-getting-started}}. This research made use of Astropy\footnote{\url{http://www.astropy.org}}, a community-developed core \py package for Astronomy \citep{astropy:2013, astropy:2018}. This research made use of data provided by Astrometry.net\footnote{\url{https://astrometry.net/use.html}}

\bibliographystyle{aa}
\bibliography{autophot}

\begin{thebibliography}{41}
\expandafter\ifx\csname natexlab\endcsname\relax\def\natexlab#1{#1}\fi

\bibitem[{{Alard} \& {Lupton}(1998)}]{alard1998}
{Alard}, C. \& {Lupton}, R.~H. 1998, \apj, 503, 325

\bibitem[{{Astropy Collaboration} {et~al.}(2018){Astropy Collaboration},
  {Price-Whelan}, {Sip{H{o}}cz}, {G{"u}nther}, {Lim}, {Crawford}, {Conseil},
  {Shupe}, {Craig}, {Dencheva}, {Ginsburg}, {Vand erPlas}, {Bradley},
  {P{'e}rez-Su{'a}rez}, {de Val-Borro}, {Aldcroft}, {Cruz}, {Robitaille},
  {Tollerud}, {Ardelean}, {Babej}, {Bach}, {Bachetti}, {Bakanov}, {Bamford},
  {Barentsen}, {Barmby}, {Baumbach}, {Berry}, {Biscani}, {Boquien}, {Bostroem},
  {Bouma}, {Brammer}, {Bray}, {Breytenbach}, {Buddelmeijer}, {Burke},
  {Calderone}, {Cano Rodr{'i}guez}, {Cara}, {Cardoso}, {Cheedella}, {Copin},
  {Corrales}, {Crichton}, {D'Avella}, {Deil}, {Depagne}, {Dietrich}, {Donath},
  {Droettboom}, {Earl}, {Erben}, {Fabbro}, {Ferreira}, {Finethy}, {Fox},
  {Garrison}, {Gibbons}, {Goldstein}, {Gommers}, {Greco}, {Greenfield},
  {Groener}, {Grollier}, {Hagen}, {Hirst}, {Homeier}, {Horton}, {Hosseinzadeh},
  {Hu}, {Hunkeler}, {Ivezi{'c}}, {Jain}, {Jenness}, {Kanarek}, {Kendrew},
  {Kern}, {Kerzendorf}, {Khvalko}, {King}, {Kirkby}, {Kulkarni}, {Kumar},
  {Lee}, {Lenz}, {Littlefair}, {Ma}, {Macleod}, {Mastropietro}, {McCully},
  {Montagnac}, {Morris}, {Mueller}, {Mumford}, {Muna}, {Murphy}, {Nelson},
  {Nguyen}, {Ninan}, {N{"o}the}, {Ogaz}, {Oh}, {Parejko}, {Parley}, {Pascual},
  {Patil}, {Patil}, {Plunkett}, {Prochaska}, {Rastogi}, {Reddy Janga},
  {Sabater}, {Sakurikar}, {Seifert}, {Sherbert}, {Sherwood-Taylor}, {Shih},
  {Sick}, {Silbiger}, {Singanamalla}, {Singer}, {Sladen}, {Sooley},
  {Sornarajah}, {Streicher}, {Teuben}, {Thomas}, {Tremblay}, {Turner},
  {Terr{'o}n}, {van Kerkwijk}, {de la Vega}, {Watkins}, {Weaver}, {Whitmore},
  {Woillez}, {Zabalza}, \& {Astropy Contributors}}]{astropy:2018}
{Astropy Collaboration}, {Price-Whelan}, A.~M., {Sip{H{o}}cz}, B.~M., {et~al.}
  2018, aj, 156, 123

\bibitem[{{Astropy Collaboration} {et~al.}(2013){Astropy Collaboration},
  {Robitaille}, {Tollerud}, {Greenfield}, {Droettboom}, {Bray}, {Aldcroft},
  {Davis}, {Ginsburg}, {Price-Whelan}, {Kerzendorf}, {Conley}, {Crighton},
  {Barbary}, {Muna}, {Ferguson}, {Grollier}, {Parikh}, {Nair}, {Unther},
  {Deil}, {Woillez}, {Conseil}, {Kramer}, {Turner}, {Singer}, {Fox}, {Weaver},
  {Zabalza}, {Edwards}, {Azalee Bostroem}, {Burke}, {Casey}, {Crawford},
  {Dencheva}, {Ely}, {Jenness}, {Labrie}, {Lim}, {Pierfederici}, {Pontzen},
  {Ptak}, {Refsdal}, {Servillat}, \& {Streicher}}]{astropy:2013}
{Astropy Collaboration}, {Robitaille}, T.~P., {Tollerud}, E.~J., {et~al.} 2013,
  \aap, 558, A33

\bibitem[{{Becker}(2015)}]{HOTPANTS_2015}
{Becker}, A. 2015, {HOTPANTS: High Order Transform of PSF ANd Template
  Subtraction}

\bibitem[{Beroiz(2019)}]{Beroiz_2019}
Beroiz, M.~I. 2019, Astrophysics Source Code Library

\bibitem[{{Bertin} \& {Arnouts}(1996)}]{Bertin1996}
{Bertin}, E. \& {Arnouts}, S. 1996, \aaps, 117, 393

\bibitem[{{Brennan} {et~al.}(2021{\natexlab{a}}){Brennan}, {Fraser},
  {Johansson}, {Pastorello}, {Kotak}, {Stevance}, {Chen}, {Eldridge}, {Bose},
  {Brown}, {Callis}, {Cartier}, {Dennefeld}, {Dong}, {Duffy}, {Elias-Rosa},
  {Hosseinzadeh}, {Hsiao}, {Kuncarayakti}, {Martin-Carrillo}, {Monard},
  {Nyholm}, {Pignata}, {Sand}, {Shappee}, {Smartt}, {Tucker}, {Wyrzykowski},
  {Abbot}, {Benetti}, {Blondin}, {Chen}, {Bento}, {Delgado}, {Galbany},
  {Gromadzki}, {Guti{\'e}rrez}, {Hanlon}, {Harrison}, {Hiramatsu}, {Hodgkin},
  {Holoien}, {Howell}, {Inserra}, {Kankare}, {Kozlowski}, {Maguire},
  {M{\"u}ller-Bravo}, {McCully}, {Meintjes}, {Morrell}, {Nicholl}, {O'Neill},
  {Pietrukowicz}, {Poleski}, {Prieto}, {Rau}, {Reichart}, {Schweyer},
  {Shahbandeh}, {Skowron}, {Sollerman}, {Sosz{\'n}yski}, {Stritzinger},
  {Szyma{\'n}ski}, {Tartaglia}, {Udalski}, {Ulaczyk}, {Young}, {van Leeuwen},
  \& {van Soelen}}]{brennan_2021a}
{Brennan}, S.~J., {Fraser}, M., {Johansson}, J., {et~al.} 2021{\natexlab{a}},
  arXiv e-prints, arXiv:2102.09572

\bibitem[{{Brennan} {et~al.}(2021{\natexlab{b}}){Brennan}, {Fraser},
  {Johansson}, {Pastorello}, {Kotak}, {Stevance}, {Chen}, {Eldridge}, {Bose},
  {Brown}, {Callis}, {Cartier}, {Dennefeld}, {Dong}, {Duffy}, {Elias-Rosa},
  {Hosseinzadeh}, {Hsiao}, {Kuncarayakti}, {Martin-Carrillo}, {Monard},
  {Pignata}, {Sand}, {Shappee}, {Smartt}, {Tucker}, {Wyrzykowski}, {Abbot},
  {Benetti}, {Blondin}, {Chen}, {Bento}, {Delgado}, {Galbany}, {Gromadzki},
  {Guti{\'e}rrez}, {Hanlon}, {Harrison}, {Hiramatsu}, {Hodgkin}, {Holoien},
  {Howell}, {Inserra}, {Kankare}, {Kozlowski}, {Maguire}, {M{\"u}ller-Bravo},
  {McCully}, {Meintjes}, {Morrell}, {Nicholl}, {O'Neill}, {Pietrukowicz},
  {Poleski}, {Prieto}, {Rau}, {Reichart}, {Schweyer}, {Shahbandeh}, {Skowron},
  {Sollerman}, {Sosz{\'n}yski}, {Stritzinger}, {Szyma{\'n}ski}, {Tartaglia},
  {Udalski}, {Ulaczyk}, {Young}, {van Leeuwen}, \& {van
  Soelen}}]{brennan_2021b}
{Brennan}, S.~J., {Fraser}, M., {Johansson}, J., {et~al.} 2021{\natexlab{b}},
  arXiv e-prints, arXiv:2102.09576

\bibitem[{{Chen} {et~al.}(2021){Chen}, {Brennan}, {Wesson}, {Fraser},
  {Schweyer}, {Inserra}, {Schulze}, {Nicholl}, {Anderson}, {Hsiao},
  {Jerkstrand}, {Kankare}, {Kool}, {Kravtsov}, {Kuncarayakti}, {Leloudas},
  {Li}, {Matsuura}, {Pursiainen}, {Roy}, {Ruiter}, {Schady}, {Seitenzahl},
  {Sollerman}, {Tartaglia}, {Wang}, {Yates}, {Yang}, {Baade}, {Carini},
  {Gal-Yam}, {Galbany}, {Gonzalez-Gaitan}, {Gromadzki}, {Gutierrez}, {Kotak},
  {Maguire}, {Mazzali}, {Mueller-Bravo}, {Paraskeva}, {Pessi}, {Pignata},
  {Rau}, \& {Young}}]{chen_2021}
{Chen}, T.~W., {Brennan}, S.~J., {Wesson}, R., {et~al.} 2021, arXiv e-prints,
  arXiv:2109.07942

\bibitem[{Craig {et~al.}(2017)Craig, Crawford, Seifert, Robitaille, Sip{\H
  o}cz, Walawender, Vin{\'{\i}}cius, Ninan, Droettboom, Youn, Tollerud, Bray,
  Walker, Janga, Stotts, G{\"u}nther, Rol, Bach, Bradley, Deil, Price-Whelan,
  Barbary, Horton, Schoenell, Heidt, Gasdia, Nelson, \&
  Streicher}]{matt_craig_2017}
Craig, M., Crawford, S., Seifert, M., {et~al.} 2017, astropy/ccdproc:
  v1.3.0.post1

\bibitem[{Foreman-Mackey {et~al.}(2013)Foreman-Mackey, Hogg, Lang, \&
  Goodman}]{Foreman_Mackey_2013}
Foreman-Mackey, D., Hogg, D.~W., Lang, D., \& Goodman, J. 2013, Publications of
  the Astronomical Society of the Pacific, 125, 306–312

\bibitem[{{Fraser} {et~al.}(2021){Fraser}, {Stritzinger}, {Brennan},
  {Pastorello}, {Cai}, {Piro}, {Ashall}, {Brown}, {Burns}, {Elias-Rosa},
  {Kotak}, {Filippenko}, {Galbany}, {Hsiao}, {Jha}, {Reguitti}, {Zhang},
  {Moran}, {Morrell}, {Shappee}, {Tomasella}, {Anderson}, {Barna}, {Ochner},
  {Phillips}, {Tucker}, {Wang}, {Baron}, {Benetti}, {Bersten}, {Brink},
  {Camacho-Neves}, {Davis}, {Dettman}, {Folatelli}, {Gutierrez}, {Hoflich},
  {Holoien}, {Kankare}, {Kumar}, {Lu}, {Mazzali}, {Taubenberger}, {Tinyanont},
  {Kuncarayakti}, {Kwok}, {Shahbandeh}, {Suntzeff}, {Yan}, {Yang}, \&
  {Zheng}}]{Fraser_2021}
{Fraser}, M., {Stritzinger}, M.~D., {Brennan}, S.~J., {et~al.} 2021, arXiv
  e-prints, arXiv:2108.07278

\bibitem[{Harbeck {et~al.}(2018)Harbeck, McCully, Pickles, Volgenau, Conway, \&
  Taylor}]{harbeck2018}
Harbeck, D.-R., McCully, C., Pickles, A., {et~al.} 2018, Long-Term Monitoring
  of Throughput in Las Cumbres Observatory's Fleet of Telescopes

\bibitem[{{Hayes} \& {Latham}(1975)}]{Hayes1975}
{Hayes}, D.~S. \& {Latham}, D.~W. 1975, \apj, 197, 593

\bibitem[{{Heasley}(1999)}]{Heasley1999}
{Heasley}, J.~N. 1999, in Astronomical Society of the Pacific Conference
  Series, Vol. 189, Precision CCD Photometry, ed. E.~R. {Craine}, D.~L.
  {Crawford}, \& R.~A. {Tucker}, 56

\bibitem[{Howell(2006)}]{howell_2006}
Howell, S.~B. 2006, Handbook of CCD Astronomy, 2nd edn., Cambridge Observing
  Handbooks for Research Astronomers (Cambridge University Press)

\bibitem[{Inn \& Tanaka(1953)}]{Inn_53}
Inn, E. C.~Y. \& Tanaka, Y. 1953, J. Opt. Soc. Am., 43, 870

\bibitem[{Kashyap {et~al.}(2010)Kashyap, van Dyk, Connors, Freeman,
  Siemiginowska, Xu, \& Zezas}]{Kashyap_2010}
Kashyap, V.~L., van Dyk, D.~A., Connors, A., {et~al.} 2010, The Astrophysical
  Journal, 719, 900–914

\bibitem[{{Landolt}(1992)}]{Landolt1992}
{Landolt}, A.~U. 1992, \aj, 104, 340

\bibitem[{Lang {et~al.}(2010)Lang, Hogg, Mierle, Blanton, \&
  Roweis}]{Lang_2010}
Lang, D., Hogg, D.~W., Mierle, K., Blanton, M., \& Roweis, S. 2010, The
  Astronomical Journal, 139, 1782–1800

\bibitem[{Lauer(1999)}]{Lauer99}
Lauer, T.~R. 1999, Publications of the Astronomical Society of the Pacific,
  111, 1434

\bibitem[{Masci(2011)}]{Masci2011ComputingFU}
Masci, F. 2011, Computing flux upper-limits for non-detections

\bibitem[{Massey \& Davis(1992)}]{Massey1992}
Massey, P. \& Davis, L. 1992, Pdf, 112, 211

\bibitem[{{McCully} \& {Tewes}(2019)}]{astroscrappy_2019}
{McCully}, C. \& {Tewes}, M. 2019, {Astro-SCRAPPY: Speedy Cosmic Ray
  Annihilation Package in Python}

\bibitem[{{Merlin} {et~al.}(2019){Merlin}, {Pilo}, {Fontana}, {Castellano},
  {Paris}, {Roscani}, {Santini}, \& {Torelli}}]{Merlin2019}
{Merlin}, E., {Pilo}, S., {Fontana}, A., {et~al.} 2019, \aap, 622, A169

\bibitem[{{Moffat}(1969)}]{Moffat_1969}
{Moffat}, A.~F.~J. 1969, \aap, 3, 455

\bibitem[{Mommert(2017)}]{Mommert_2017}
Mommert, M. 2017, Astronomy and Computing, 18, 47–53

\bibitem[{{Mortara} \& {Fowler}(1981)}]{Mortara_1981}
{Mortara}, L. \& {Fowler}, A. 1981, in Society of Photo-Optical Instrumentation
  Engineers (SPIE) Conference Series, Vol. 290, Society of Photo-Optical
  Instrumentation Engineers (SPIE) Conference Series, 28

\bibitem[{Patat {et~al.}(2011)Patat, Moehler, O’Brien, Pompei, Bensby,
  Carraro, de~Ugarte~Postigo, Fox, Gavignaud, James, \& et~al.}]{Patat_2011}
Patat, F., Moehler, S., O’Brien, K., {et~al.} 2011, Astronomy \&
  Astrophysics, 527, A91

\bibitem[{Perley {et~al.}(2018)Perley, Mazzali, Yan, Cenko, Gezari, Taggart,
  Blagorodnova, Fremling, Mockler, Singh, \& et~al.}]{perley_2018}
Perley, D.~A., Mazzali, P.~A., Yan, L., {et~al.} 2018, Monthly Notices of the
  Royal Astronomical Society, 484, 1031–1049

\bibitem[{{Prentice} {et~al.}(2018{\natexlab{a}}){Prentice}, {Ashall},
  {Mazzali}, {Zhang}, {James}, {Wang}, {Vink{\'o}}, {Percival}, {Short},
  {Piascik}, {Huang}, {Mo}, {Rui}, {Wang}, {Xiang}, {Xin}, {Yi}, {Yu}, {Zhai},
  {Zhang}, {Hosseinzadeh}, {Howell}, {McCully}, {Valenti}, {Cseh}, {Hanyecz},
  {Kriskovics}, {P{\'a}l}, {S{\'a}rneczky}, {S{\'o}dor}, {Szak{\'a}ts},
  {Sz{\'e}kely}, {Varga-Vereb{\'e}lyi}, {Vida}, {Bradac}, {Reichart}, {Sand},
  \& {Tartaglia}}]{Prentice_2018a}
{Prentice}, S.~J., {Ashall}, C., {Mazzali}, P.~A., {et~al.} 2018{\natexlab{a}},
  \mnras, 478, 4162

\bibitem[{{Prentice} {et~al.}(2018{\natexlab{b}}){Prentice}, {Maguire},
  {Smartt}, {Magee}, {Schady}, {Sim}, {Chen}, {Clark}, {Colin}, {Fulton},
  {McBrien}, {O'Neill}, {Smith}, {Ashall}, {Chambers}, {Denneau}, {Flewelling},
  {Heinze}, {Holoien}, {Huber}, {Kochanek}, {Mazzali}, {Prieto}, {Rest},
  {Shappee}, {Stalder}, {Stanek}, {Stritzinger}, {Thompson}, \&
  {Tonry}}]{Prentice_2018b}
{Prentice}, S.~J., {Maguire}, K., {Smartt}, S.~J., {et~al.} 2018{\natexlab{b}},
  \apjl, 865, L3

\bibitem[{{Prentice} \& {Mazzali}(2017)}]{Prentice_2017}
{Prentice}, S.~J. \& {Mazzali}, P.~A. 2017, \mnras, 469, 2672

\bibitem[{{Science Software Branch at STScI}(2012)}]{pyraf2012}
{Science Software Branch at STScI}. 2012, {PyRAF: Python alternative for IRAF}

\bibitem[{{Stetson}(1987)}]{DAOPHOT1987}
{Stetson}, P.~B. 1987, \pasp, 99, 191

\bibitem[{Stritzinger {et~al.}(2002)Stritzinger, Hamuy, Suntzeff, Smith,
  Phillips, Maza, Strolger, Antezana, González, Wischnjewsky, \&
  et~al.}]{Stritzinger2002}
Stritzinger, M., Hamuy, M., Suntzeff, N.~B., {et~al.} 2002, The Astronomical
  Journal, 124, 2100–2117

\bibitem[{{Tody}(1986)}]{toby_iraf1986}
{Tody}, D. 1986, in Society of Photo-Optical Instrumentation Engineers (SPIE)
  Conference Series, Vol. 627, Instrumentation in astronomy VI, ed. D.~L.
  {Crawford}, 733

\bibitem[{{Tody}(1993)}]{toby_iraf1993}
{Tody}, D. 1993, in Astronomical Society of the Pacific Conference Series,
  Vol.~52, Astronomical Data Analysis Software and Systems II, ed. R.~J.
  {Hanisch}, R.~J.~V. {Brissenden}, \& J.~{Barnes}, 173

\bibitem[{{van Dokkum} {et~al.}(2012){van Dokkum}, {Bloom}, \&
  {Tewes}}]{vanDokkum_2012}
{van Dokkum}, P.~G., {Bloom}, J., \& {Tewes}, M. 2012, {L.A.Cosmic: Laplacian
  Cosmic Ray Identification}

\bibitem[{Wildey(1992)}]{Wildey_1992}
Wildey, R.~L. 1992, Publications of the Astronomical Society of the Pacific,
  104, 285

\bibitem[{Zackay {et~al.}(2016)Zackay, Ofek, \& Gal-Yam}]{Zackay_2016}
Zackay, B., Ofek, E.~O., \& Gal-Yam, A. 2016, The Astrophysical Journal, 830,
  27

\end{thebibliography}

\end{document}